\documentclass[aps, superscriptaddress]{revtex4-1}

\usepackage[dvipdfmx]{graphicx}
\usepackage{amsmath}
\usepackage{amssymb}
\usepackage{bm}
\usepackage{color}

\begin{document}
	\title{Phase dynamics of noise-induced coherent oscillations in excitable systems}
	\author{Jinjie Zhu}
	\email{zhu.j.ag@m.titech.ac.jp}
	\affiliation{School of Mechanical Engineering, Nanjing University of Science and Technology, Nanjing 210094, China}
	\affiliation{Department of Systems and Control Engineering, Tokyo Institute of Technology, Tokyo 152-8552, Japan}
	\author{Yuzuru Kato}
	\affiliation{Department of Complex and Intelligent Systems, Future University Hakodate, Hokkaido 041-8655, Japan}
	\author{Hiroya Nakao}
	\email{nakao@sc.e.titech.ac.jp}
	\affiliation{Department of Systems and Control Engineering, Tokyo Institute of Technology, Tokyo 152-8552, Japan}
	\date{\today}
	
	\begin{abstract}
		Noise can induce coherent oscillations in excitable systems without periodic orbits. Here, we establish a method to derive a hybrid system approximating the noise-induced coherent oscillations in excitable systems and further perform phase reduction of the hybrid system to derive an effective, dimensionality-reduced phase equation. We apply the reduced phase model to a periodically forced excitable system and two-coupled excitable systems, both undergoing noise-induced oscillations. The reduced phase model can quantitatively predict the entrainment of a single system to the periodic force and the mutual synchronization of two coupled systems, including the phase slipping behavior due to noise, as verified by Monte Carlo simulations. The derived phase model gives a simple and efficient description of noise-induced oscillations and can be applied to the analysis of more general cases.
	\end{abstract}
	
	\maketitle
	
	{\it Introduction.-}
	Noise is ubiquitous in nature and is generally considered to hinder ordered behaviors of systems. However, counterintuitive phenomena in which noise brings order have also been revealed and aroused much attention in diverse fields. Indeed, noise can facilitate the formation of self-organized structures in nonequilibrium systems in physics, chemistry, and biology~\cite{Prigogine2017book,Horsthemke1984book,Bressloff2013RMP}. For example, synchronization of nonlinear oscillators usually requires mutual coupling, but common or correlated noise applied on them can induce synchronization even when the oscillators are uncoupled~\cite{Zhou2002PRL,Teramae2004PRL,Nakao2007PRL}. In nonlinear systems, stochastic trajectories under the effect of noise are not necessarily blurred versions of the deterministic trajectories~\cite{Alexandrov2021PR}. Noise may induce coherent trajectories that do not exist without noise, e.g., in stochastic resonance, coherence resonance, noise-induced synchronization, spatiotemporal patterns, etc. \cite{Lindner2004PR}. In particular, in excitable systems, even if no periodic orbit exists in the absence of noise, coherent oscillations can still occur when noise with appropriate intensity is applied, which resemble deterministic limit-cycle oscillations.
	
	Phase reduction is a powerful tool for reducing the dimensionality of limit-cycling systems under weak perturbations \cite{Kuramoto1984,Nakao2016}. Due to its simplicity and efficiency, this approach has been widely employed in analyzing various systems of coupled oscillators and also generalized to nonconventional systems such as delay-induced oscillations~\cite{Kotani2012PRL,Kotani2020PRR}, reaction-diffusion systems~\cite{Nakao2014PRX}, stochastic limit-cycle oscillators~\cite{Teramae2009PRL,Nakao2007PRL,Lai2011PRL}, hybrid oscillators~\cite{Shirasaka2017PRE,Park2018EJAM}, relaxation oscillators~\cite{Izhikevich2000SIAMJAM,Zhu2020PRE}, quantum nonlinear oscillators~\cite{Kato2019PRR,Kato2020PRE}, etc. The phase reduction relies on the notion of the asymptotic phase \cite{Winfree1980,Kuramoto1984,Nakao2016} of the limit cycle to characterize the dominant dynamical behaviors. However, it is not an easy task to establish a phase reduction theory for noise-induced coherent excitable systems due to the lack of a reference periodic orbit that characterizes the coherent oscillations. Efforts have so far been made mainly to develop a phenomenological phase model based on numerical simulations and data processing~\cite{Han1999PRL,Kralemann2008PRE,Schwabedal2010PRE}, or to define the stochastic version of the asymptotic phase and amplitude by solving the eigenvalue problem of the backward Kolmogorov operator for stochastic oscillators~\cite{Thomas2014PRL,Perez-Cervera2021PRL,Kato2021Mathematics}.
	
	In this Letter, we construct a quantitative phase reduction theory for noise-induced coherent excitable systems
	by (i) finding a reference orbit which plays the role of the limit cycle in deterministic oscillatory cases; (ii) establishing an approximate hybrid system for calculating the phase sensitivity function; and (iii) constructing an effective phase equation and applying it to the analysis of periodically forced or mutually coupled oscillators.
	
\begin{figure}
		\centering
		\includegraphics[width=0.6\textwidth]{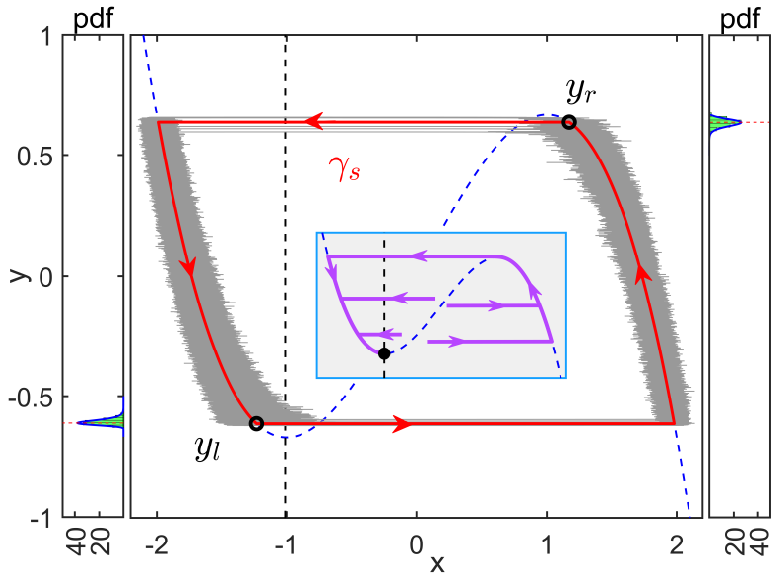}
		\caption{Prediction of a stochastic periodic orbit $\gamma_s$ (red bold curve) approximating the stochastic trajectories (gray) obtained by MC simulations of the excitable FHN system exhibiting SISR oscillations. The black circles are two transition positions on the left and right branches obtained via DMC. The dashed curves are $x$ (blue) and $y$ (black) nullclines, respectively. Left and right panels show the distributions of the transition position on the left and right branches obtained by MC simulations (green bars) and FPTD (blue curves). Inset: Deterministic dynamics of the excitable FHN system (\ref{eq:1}) without noise. The black dot is a stable fixed point ($x_0$, $y_0$). }
		\label{fig:1}       
	\end{figure}

	
	{\it Phase reduction of noise-induced coherent systems.-} We consider a FitzHugh-Nagumo (FHN) system perturbed by noise applied on the fast variable as a typical example:
	\begin{equation}
		\begin{split}
			\varepsilon \dot x &= f(x)-y + \sqrt{D_{\nu}}\nu(t),\\
			\dot{y} &= x+a,
		\end{split}
		\label{eq:1}
	\end{equation}
	where $x$ and $y$ represent the fast membrane potential and slow recovery variable, respectively, and $y=f(x)=x-\frac{x^3}{3}$ is the nullcline of $x$. The Gaussian white noise $\nu(t)$ satisfies $\left<\nu(t)\right> = 0$ and $\left<\nu(t)\nu(\tau)\right> = \delta(t-\tau)$, and $D_{\nu}$ represents its intensity. The timescale separation parameter $\varepsilon$ and the bifurcation parameter $a$ are fixed as $\varepsilon=0.0001$ and $a=1.01$. Without noise, the system (\ref{eq:1}) has a globally stable fixed point ($x_0$, $y_0$), where $x_0=-a$ and $y_0=\frac{a^3}{3}-a$ (see the inset in Fig.~\ref{fig:1}). When noise is applied, the system (\ref{eq:1}) can exhibit noisy but coherent oscillations due to large timescale separation, called self-induced stochastic resonance (SISR) \cite{Muratov2005PD}. Figure \ref{fig:1} illustrates the SISR oscillations observed at noise intensity $D_{\nu}=0.01$. We refer to the system (\ref{eq:1}) as the SISR oscillator hereafter.
	
	To determine the reference periodic orbit characterizing the coherent oscillations, we apply the distance matching condition (DMC) that we developed in Ref.~\cite{Zhu2021PRR} to calculate the transition positions (see Supplemental Material (SM) \cite{supplement} for details). The transition position $y_l$ of the stochastic trajectory started from the initial state $y_0$ on the left branch can be determined from the following DMC:
	\begin{equation}
		\int_{y_0}^{y_{l}} \frac{S(y)}{\varepsilon \left[f_l^{-1}(y)+a\right] \, T_e(y)}\,dy=S(y_{l}),
		\label{eq:2}
	\end{equation}
	where $f_l^{-1}(y)$ is the value of $x$ on the left branch at $y$ and $T_e(y)$ is the mean first passage time. The left-hand side of Eq. (\ref{eq:2}) represents the accumulated effect of noise on the displacement of the state away from the stable branch and $S(y)$ on the right-hand side is the distance between the middle and left branches. This condition implies that the transition occurs when the noise-induced displacement and the distance from the left to the middle branch match. As shown in Fig.~\ref{fig:1}, the transition positions on the left and right branches predicted by Eq. (\ref{eq:2}) are in good agreement with the Monte Carlo (MC) simulations. It is found that the transition position can also be predicted by considering the first passage time distribution (FPTD) \cite{Gardiner1985,Lim2010JCN,Li2019PRE}, which can be calculated for the left branch (and similarly for the right branch) as~\cite{supplement}
	\begin{equation}
		\rho_{l}(y)=\frac{\exp\left(-\int^{y}\frac{1}{\varepsilon \left[f_{l}^{-1}(y^\prime)+a\right] T_e(y^\prime)}\mathrm{d}y^\prime\right)}{\varepsilon \left|f_{l}^{-1}(y)+a\right| T_e(y)}.
		\label{eq:3}
	\end{equation}
	The peak of FPTD agrees well with the transition position predicted from DMC on each branch as shown in Fig.~\ref{fig:1}.
	
	It is noted that the stochastic oscillations caused by SISR possess a well-defined orbit and almost keep a deterministic period. These features pave the way for applying the phase reduction approach to this system. This stochastic periodic orbit is completely different from the deterministic orbit in the oscillatory regime of the system and cannot be approximated by some limiting process of the latter. In particular, the transition from one branch to the other happens before reaching the tips of the $x$ nullcline. In order to simplify our analysis, we fix the noise intensity $D_{\nu}=0.01$ in what follows. However, we note that the SISR phenomenon can also be induced at different noise intensities as shown in Ref.~\cite{Zhu2021PRR} and the proposed reduction method is generally applicable to a wide range of parameters.
	
	Considering the fast-slow characteristics of the stochastic periodic orbit, we approximate the SISR oscillator by using the following hybrid (piecewise-continuous) dynamical system:
	\begin{equation}
		\begin{split}
			&\dot {\bm X} = {\bm F}({\bm X}), \text{if} ~{\bm X}\notin{\bm \Pi_i},\\
			&{\bm X}(t+0) = {\bm \Phi_i}({\bm X}(t)), \text{if} ~{\bm X}\in{\bm \Pi_i}, i=l,r.
		\end{split}
		\label{eq:4}
	\end{equation}
	Here, ${\bm X} = (x, y)$, ${\bm F}({\bm X})$ is the deterministic vector field of the system (\ref{eq:1}), ${\bm \Pi_i}$ are switching surfaces on the left and right branches, and ${\bm \Phi}_i$ are transition functions. That is, we approximate the slow stochastic dynamics along the left and right branches by the deterministic orbit of the original system and the fast dynamics between the branches by instantaneous discontinuous transitions. The transition functions and the switching surfaces can be calculated as ${\bm \Phi_l}({\bm X}) = \left[2 \cos(\varphi), y\right]^\top$, ${\bm \Phi_r}({\bm X}) = \left[2 \cos\left(\varphi+\frac{2\pi}{3}\right), y\right]^\top$, and ${\bm \Pi_i}=\left\{{\bm X}|L({\bm X})=y_{i} \right\}$, where $L({\bm X})=y$, $\varphi=\frac{1}{3} \arccos\left(-\frac{3}{2}y\right)$, and $y_{l}$ and $y_{r}$ are the transition positions on the left and right branches obtained by DMC \cite{supplement}. This system has a stable, piecewise-continuous limit cycle, denoted as $\gamma_s$, of frequency $\omega_h$, where $\omega_h$ is nearly equal to the average frequency of the stochastic oscillations.
	
\begin{figure}
		\centering
		\includegraphics[width=0.6\textwidth]{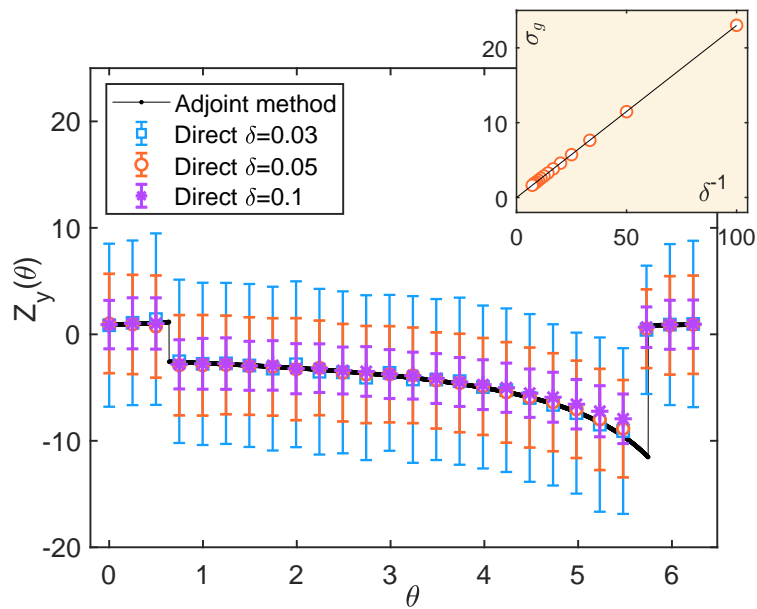}
		\caption{Phase sensitivity function [$y$ component $Z_y(\theta)$]. The black curve shows the theoretical result obtained by solving the adjoint equation of the hybrid system (\ref{eq:4}). Symbols are the results obtained by directly applying impulses of strength $\delta$ on the SISR oscillator (\ref{eq:1}) and measuring the resulting phase differences after $T_h = 2\pi \omega_h^{-1}$~\cite{supplement}; the error bars are their standard deviations. Inset: Linear fitting of the standard deviation $\sigma_g$ versus $\delta^{-1}$ \cite{supplement}; circles are results by the direct measurement and the black line is the linear fitting with $\sigma_{g}=0.2300\delta^{-1}-0.0084$.
			\label{fig:2}
		}
	\end{figure}

	Through the above approximation, we have transformed the original stochastic system (\ref{eq:1}) to the hybrid system (\ref{eq:4}). We now apply the phase reduction method for hybrid systems \cite{Shirasaka2017PRE} to further reduce the system (\ref{eq:4}) into the phase equation of the form	${\dot \theta(t)}=\frac{\mathrm{d} \Theta({\bm X}(t))}{\mathrm{d} t}=\frac{\partial \Theta({\bm X})}{\partial {\bm X}} \cdot \bm F(\bm X)=\omega_h$, where $\omega_h$ is the frequency of Eq. (\ref{eq:4}) and $\theta(t) = \Theta({\bm X}(t))$ is the phase of the system. Here, the phase function $\Theta({\bm X})$ gives the asymptotic phase \cite{Winfree1980,Kuramoto1984,Nakao2016} of the state ${\bm X}$ within the basin of attraction of the limit cycle $\gamma_s$. When the SISR oscillator is additionally subjected to a weak perturbation $\bm P(\bm X,t)$, the first-order approximate phase dynamics is given by ${\dot \theta(t)}=\omega_h+\frac{\partial \Theta({\bm X})}{\partial {\bm X}} \cdot \bm P(\bm X,t) \approx \omega_h+ \bm Z(\theta) \cdot \bm P\left(\theta,t\right)$, where we have approximated the system state $\bm X(t)$ by the state $\bm X_0(\theta(t))$ on $\gamma_s$ sharing the same asymptotic phase, and $\bm Z(\theta)=\frac{\partial \Theta({\bm X})}{\partial {\bm X}}\big|_{{\bm X}={\bm X}_0(\theta)}$ is the phase sensitivity function of $\gamma_s$, which can be obtained via the adjoint method for hybrid limit cycles \cite{supplement}. The $y$ component of the phase sensitivity functions $Z_y(\theta)$ is illustrated in Fig.~\ref{fig:2}, which has discontinuities at the two transition positions.
	
	In the above derivation of the phase equation, we omitted the stochastic fluctuations of the SISR oscillator. To better describe the stochastic dynamical behaviors and quantitatively evaluate the accuracy of our prediction, we further incorporate the stochasticity of the system into the phase equation as an effective additive noise~\cite{Schwabedal2010PRE,Nakao2010Chaos},
	\begin{equation}
		{\dot \theta(t)}=\omega_e+\bm Z(\theta) \cdot \bm P\left(\theta,t\right)+\sqrt{D_e} \xi(t),
		\label{eq:5}
	\end{equation}
	where $\omega_e$ denotes the effective frequency of the stochastic oscillations, $\xi(t)$ is the Gaussian-white noise satisfying $\left<\xi(t)\right> = 0$ and $\left<\xi(t)\xi(\tau)\right> = \delta(t-\tau)$ , and $D_e$ represents the effective noise intensity. The effective frequency and noise intensity are evaluated by the ensemble average \cite{Nakao2010Chaos} as $\omega_e=\left<[\theta(t)-\theta(0)]/t\right>$ and $D_e=\left<\left([\theta(t)-\theta(0)]/t-\omega_e\right)^2\right>t$, respectively, by MC simulations of the original system~(\ref{eq:1}), which are obtained as $\omega_e=2.5161$ and $D_e=0.0104$. As discussed in SM \cite{supplement}, the approximate theoretical value of the effective frequency can be obtained from DMC as $\tilde{\omega_e}=\omega_h=2.5266$ and that of the effective noise intensity can be evaluated from FPTD as $\tilde{D_e}=0.0095$, which agree well with the values of $\omega_e$ and $D_e$ and quantitatively validate the hybrid system (\ref{eq:4}).
	
	We can also evaluate the effective noise intensity by direct measurement of the phase sensitivity function \cite{supplement}. The $y$ component of the phase sensitivity functions $Z_y(\theta)$ evaluated using several different perturbation intensities is shown in Fig.~\ref{fig:2}. As discussed in SM \cite{supplement}, as the perturbation intensity $\delta$ used for the measurement becomes smaller, the mean value of the measured $Z_y(\theta)$ approaches the theoretical result for infinitesimal perturbation intensity calculated by the adjoint method, while its standard deviation increases as $\sigma_g = \delta^{-1}\sqrt{2D_e t}$. From the inset of Fig.~\ref{fig:2}, the effective noise intensity is evaluated as $D_e = 0.0106$, which is also consistent with the values obtained by the other methods. In the following analysis, we fix the parameters as $\omega_e=2.5161$ and $D_e=0.0104$ in the effective phase equation (\ref{eq:5}). As we will demonstrate, the simple reduced phase equation~(\ref{eq:5}) that we have derived can accurately predict the dynamical behaviors of the SISR oscillator (\ref{eq:1}) under general weak perturbations, such as the periodic forcing and mutual coupling.
	
	
	{\it Periodic forcing.-}
	We first consider a periodically forced SISR oscillator described by
	\begin{equation}
		\begin{split}
			\varepsilon \dot x &= f(x)-y + \sqrt{D_{\nu}}\nu(t),\\
			\dot{y} &= x+a+\mu \sin(\Omega t),
		\end{split}
		\label{eq:6}
	\end{equation}
	where the forcing frequency $\Omega$ is close to the effective frequency $\omega_e$ and $\mu$ characterizes the strength of the periodic forcing, which is weak in the sense that the frequency difference is given by $\omega_e - \Omega = \mu \Delta$ where $|\mu| \ll 1$ and $\Delta$ is of $O(1)$. By applying the reduction method described above, we obtain the reduced phase equation $\dot \theta=\omega_e+\sqrt{D_e}\xi(t)+\mu Z_y(\theta)\sin(\Omega t)$. By further introducing the slow relative phase $\phi(t)=\theta(t)-\Omega t$, we obtain $\dot \phi=\omega_e-\Omega+\sqrt{D_e}\xi(t)+\mu Z_y(\phi+\Omega t)\sin(\Omega t).$ Since $\phi(t)$ varies much more slowly than $\Omega t$ because $\omega_e - \Omega$ is of $O(\mu)$, we can average this equation via the corresponding Fokker-Planck equation over one period of fast oscillation~\cite{Kuramoto1984}. This yields a further simplified phase equation
	\begin{equation}
		\dot \phi=\mu \left(\Delta+\Gamma_p(\phi)\right)+\sqrt{D_e}\xi(t),
		\label{eq:7}
	\end{equation}
	where $\Gamma_p(\phi) = \frac{1}{2\pi}\int_{0}^{2\pi} Z_y(\phi+\psi)\sin(\psi) \mathrm{d}\psi$ is a phase coupling function representing the effect of the periodic forcing on the phase dynamics.
	
\begin{figure*}
		\centering
		\includegraphics[width=0.3\textwidth]{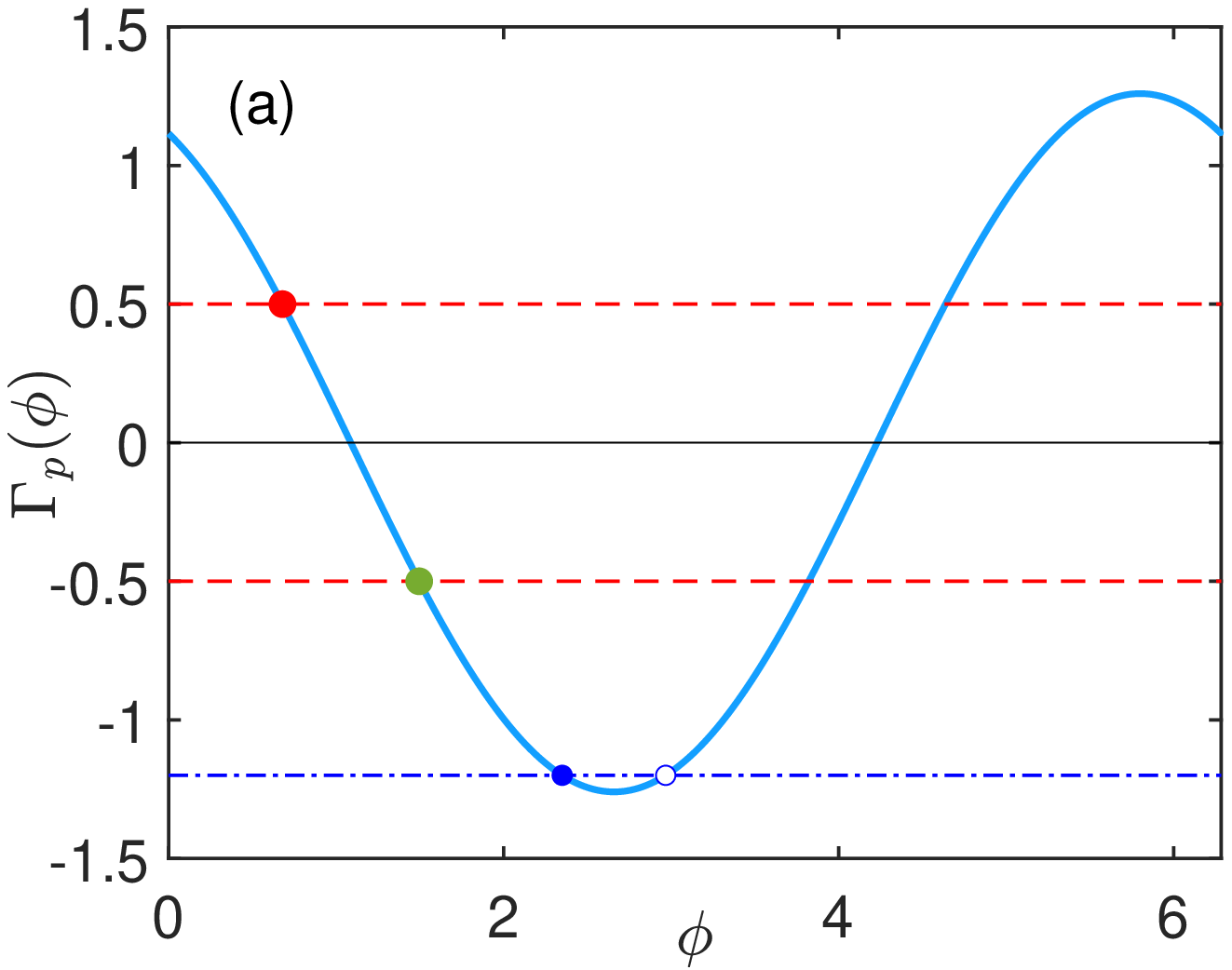}
		\includegraphics[width=0.3\textwidth]{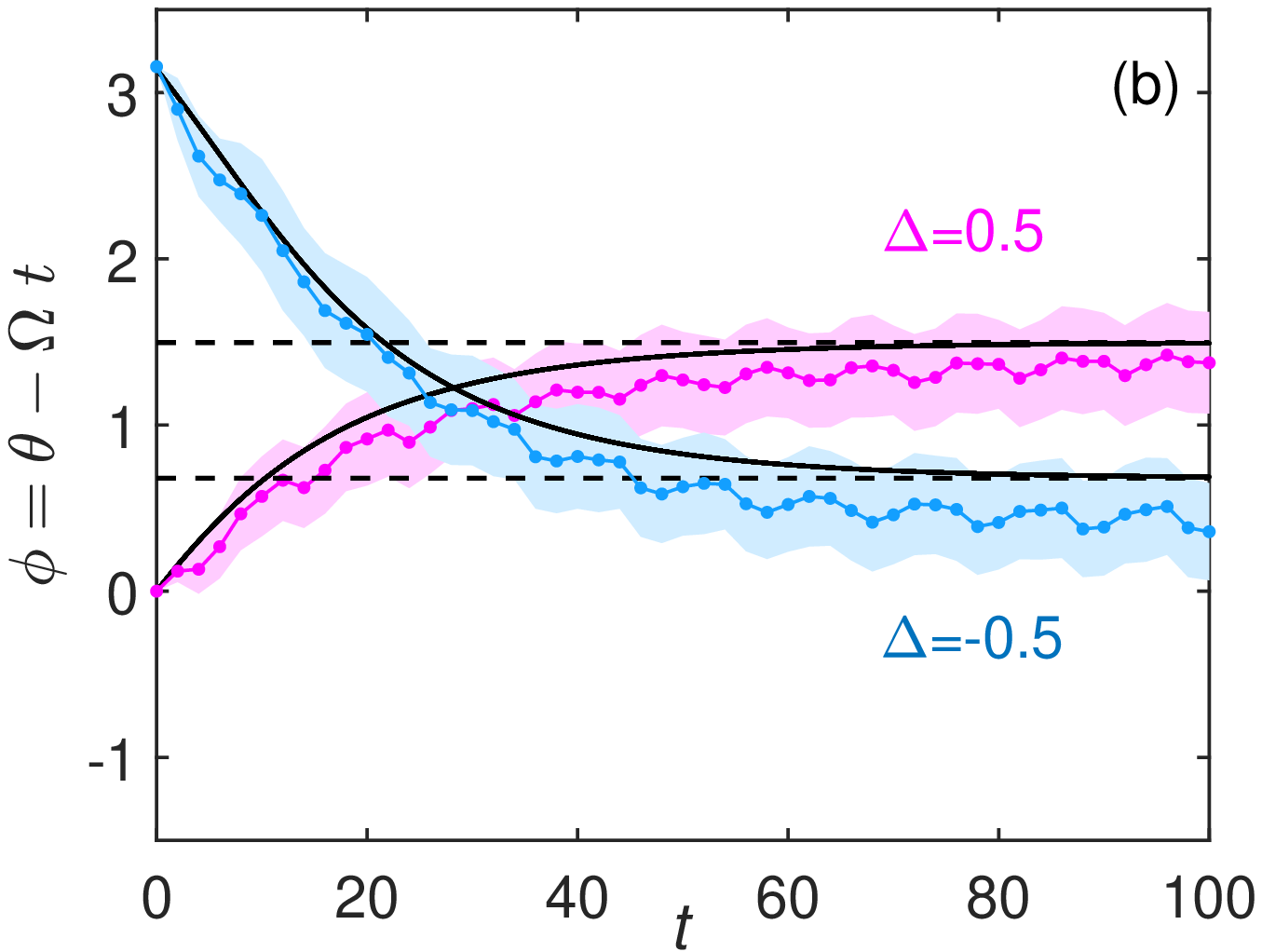}
		\includegraphics[width=0.3\textwidth]{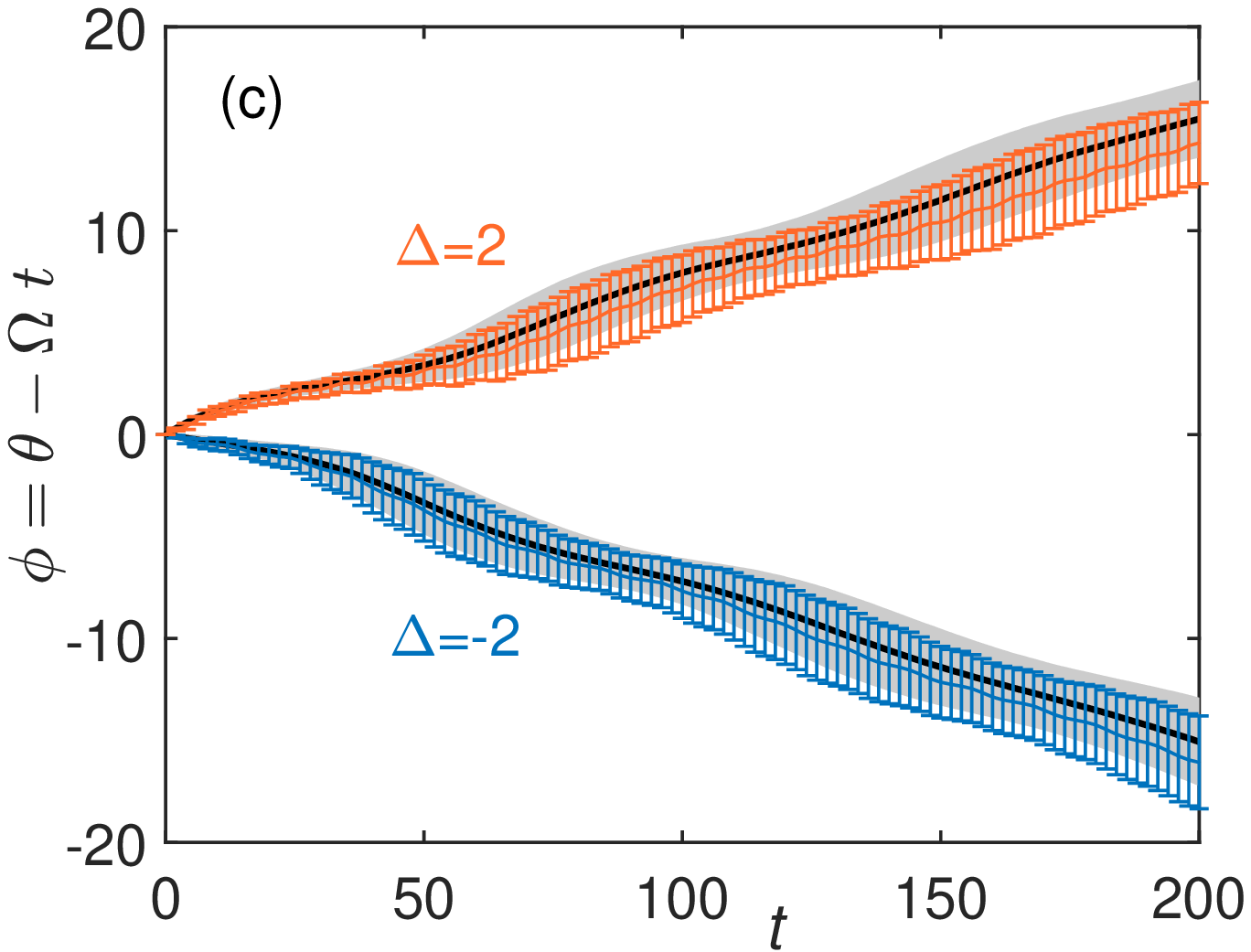}
		\includegraphics[width=0.3\textwidth]{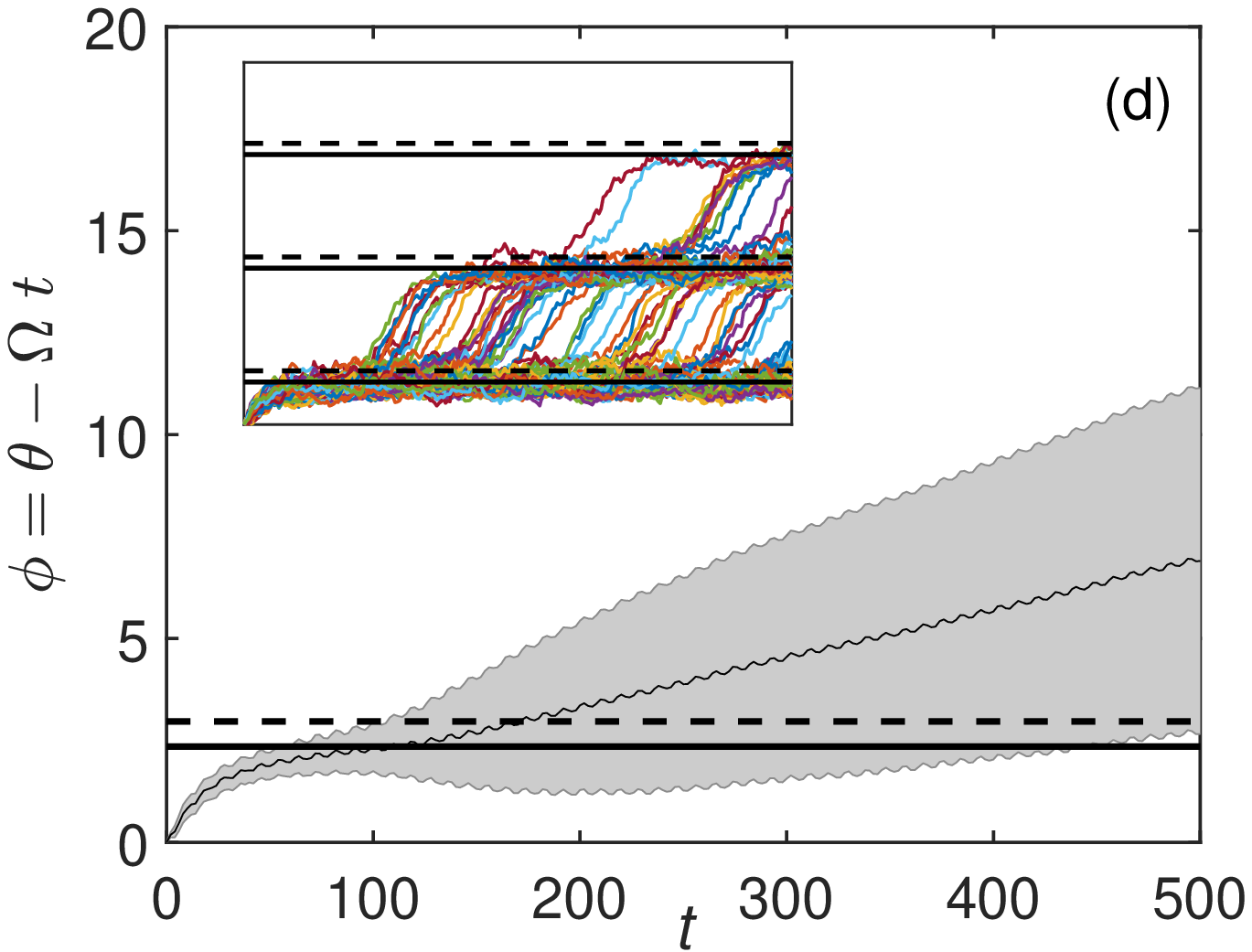}
		\includegraphics[width=0.3\textwidth]{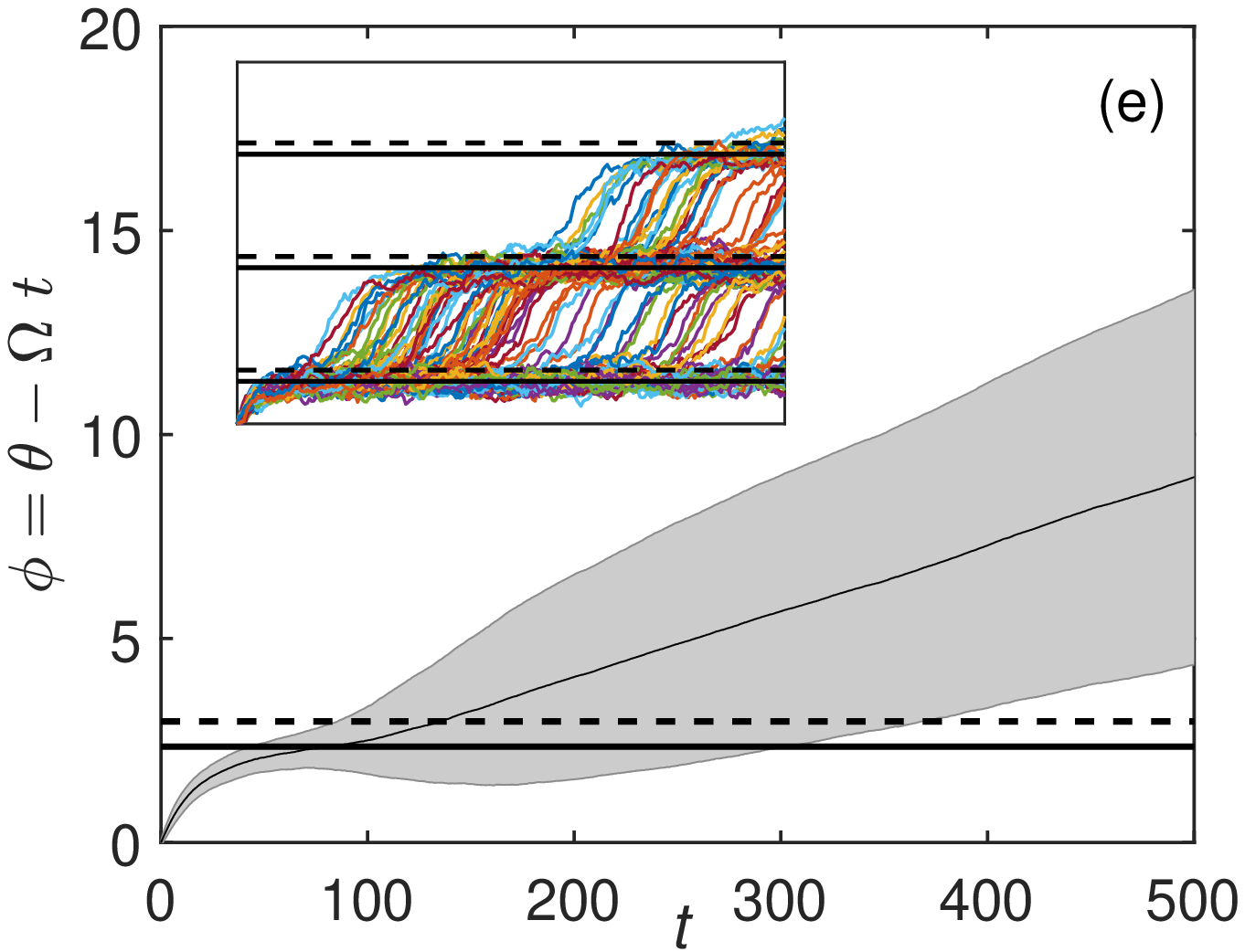}
		\includegraphics[width=0.3\textwidth]{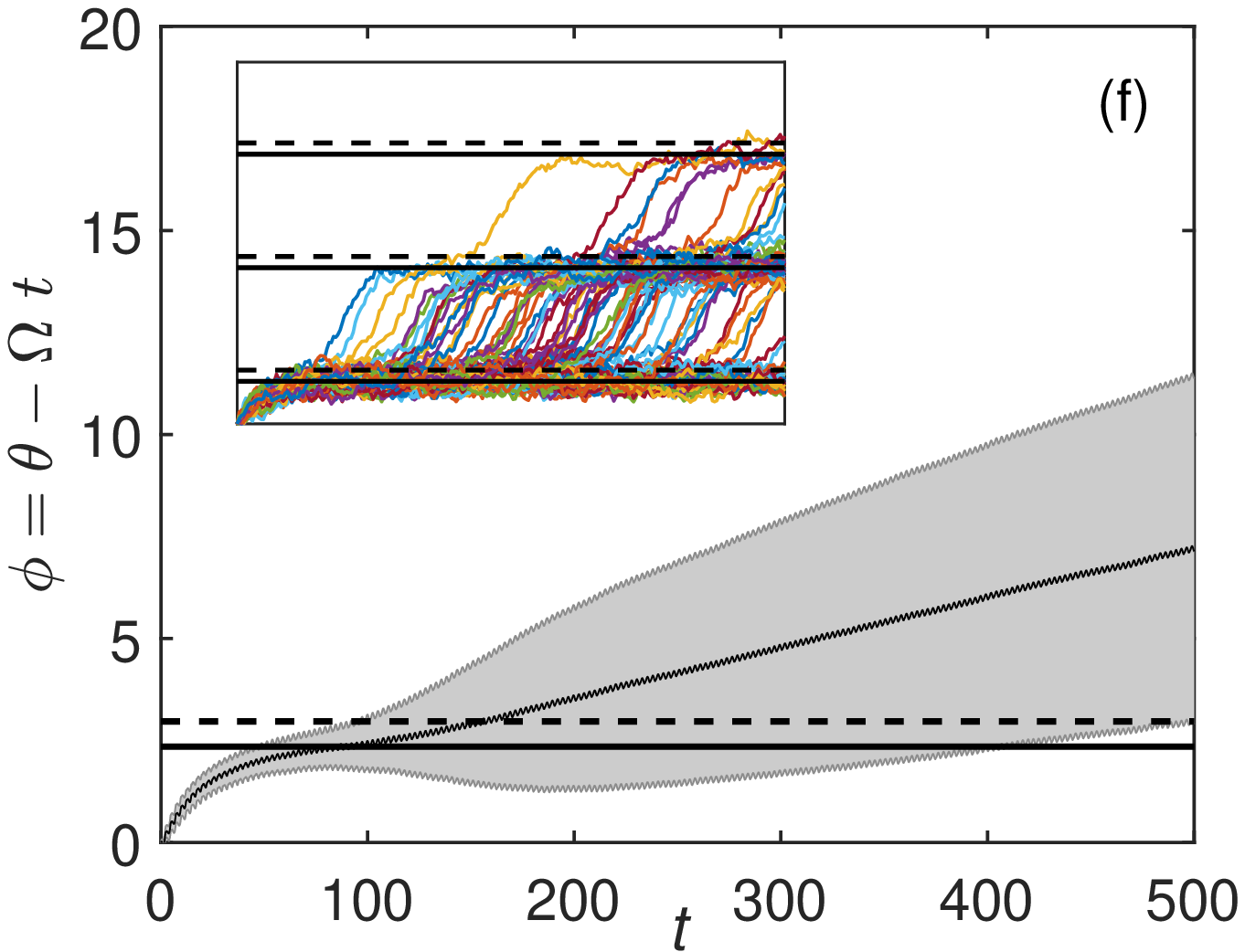}
		\caption{(a) Phase coupling function $\Gamma_p(\phi)$. Two red dashed lines are for the cases of $\Delta=0.5$ (below) and $\Delta=-0.5$ (above), respectively. The blue dot-dashed line is for the case $\Delta=1.2$. (b) Dynamics of the relative phase $\phi$ for $\Delta=0.5$ and $\Delta=-0.5$. Colored curves: MC simulations; black curves: Eq. (\ref{eq:7}) without noise. (c) Dynamics of $\phi$ for $\Delta=2$ and $\Delta=-2$. Colored curves: MC simulations; black curves: Eq. (\ref{eq:7}). (d)--(f) Dynamics of $\phi$ for $\Delta=1.2$ of the original perturbed system (\ref{eq:6}), averaged phase equation (\ref{eq:7}), and phase equation before averaging, respectively. The insets display 100 realizations of MC simulations. The solid and dashed lines represent the stable and unstable phase-locked states predicted in (a). The shaded areas and error bars represent standard deviations of the results. The coupling strength is $\mu=0.05$.}
		\label{fig:3}       
	\end{figure*}

	The phase coupling function $\Gamma_p(\phi)$ is plotted in Fig.~\ref{fig:3}(a), which is smooth despite that the phase sensitivity function is discontinuous because $\Gamma_p(\phi)$ is a convolution of $Z_y$ with a smooth function. The solution to $\Gamma_p(\phi)=-\Delta$ gives the phase-locked state and it is linearly stable (unstable) when $\Gamma^{\prime}_p(\phi)<0$ [$\Gamma^{\prime}_p(\phi) > 0$] if the noise is not considered. Figure~\ref{fig:3}(b) shows the time evolution of the relative phase $\phi$ converging to the stable phase-locked state for the cases with $\Delta=0.5$ and $\Delta=-0.5$. The results of MC simulations of the system (\ref{eq:6}) are in good agreement with the theoretical prediction by the noiseless phase model with the phase coupling function in Fig.~\ref{fig:3}(a). When the frequency difference is above the critical value, i.e., $|\Delta|>|\Delta_c|\approx1.2596$, the relative phase will continue to increase or decrease with time because the noiseless system does not have stable phase-locked states, and phase drift will occur \cite{Pikovsky2001book}. Figure \ref{fig:3}(c) shows that this can also be well predicted by using the reduced phase model.
	
	When $| \Delta |$ is slightly below the critical value, noise-induced phase slipping \cite{Pikovsky2001book,Nakao2016,Stankovski2012PRL,Berglund2014ArXiv} can be observed. Figure~\ref{fig:3}(d) shows the results of MC simulations of Eq. (\ref{eq:6}). The relative phase $\phi$ converges to the stable phase-locked state, but it can occasionally cross the unstable phase-locked state due to noise and exhibits phase slips. This phenomenon can also be well reproduced by the reduced phase equation (\ref{eq:7}) as shown in Fig.~\ref{fig:3}(e), plotting the results of MC simulations for the averaged phase model. The discrepancy of Fig.~\ref{fig:3}(e) from Fig.~\ref{fig:3}(d) mainly originates from the averaging approximation. By directly performing simulations of the reduced phase equation before averaging, the discrepancy from the original model can be significantly reduced as shown in Fig.~\ref{fig:3}(f).
	
\begin{figure*}
		\centering
		\includegraphics[width=0.9\textwidth]{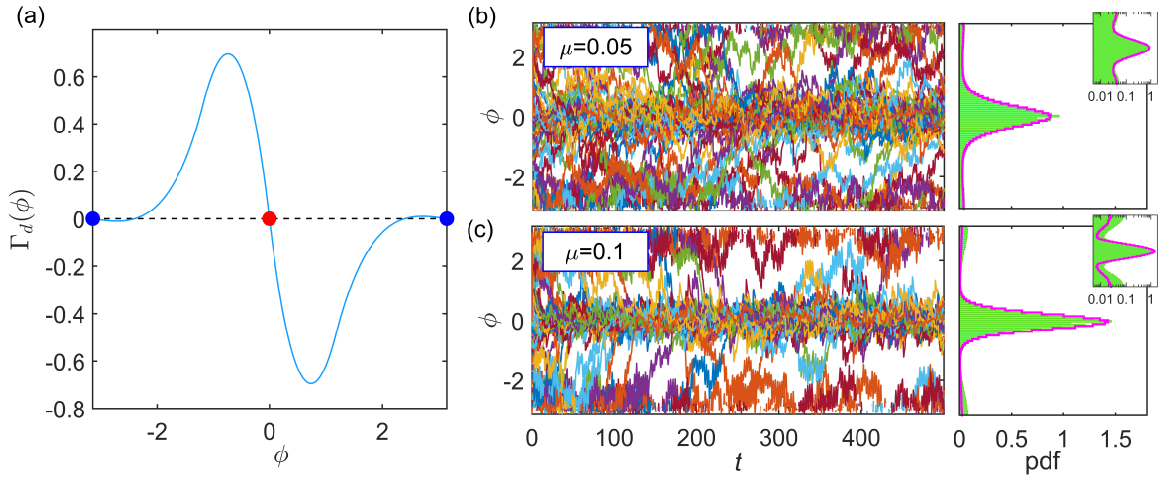}
		\caption{\label{fig:4}(a) Phase coupling function $\Gamma_d(\phi)$. The dots represent stable synchronized states ($\phi=0,\pm \pi$). (b),(c) Dynamics of the phase difference $\phi$ for two identical SISR oscillators with mutual coupling. Left panel: Time series of the phase difference for coupling strengths $\mu=0.05$ and $\mu=0.1$ from uniformly distributed initial conditions obtained by MC simulations of the system (\ref{eq:8}). Right panel: Phase-difference distribution ($400 \leq t \leq 500$). Green bars: Results of MC simulations of the system (\ref{eq:8}); pink curves: prediction of the reduced phase equation (\ref{eq:9}).
		}
	\end{figure*}

	{\it Two-coupled SISR oscillators.-}
	Next, we consider two weakly coupled identical SISR oscillators described by
	\begin{equation}
		\begin{split}
			\varepsilon \dot x_1 &= f(x_1)-y_1 + \sqrt{D_\nu}\nu_1(t),~\dot{y_1} = (x_1+a) + \mu (y_2-y_1),\\
			\varepsilon \dot x_2 &= f(x_2)-y_2 + \sqrt{D_\nu}\nu_2(t),~\dot{y_2} = (x_2+a) + \mu (y_1-y_2),
		\end{split}
		\label{eq:8}
	\end{equation}
	where $\mu$ ($0 < \mu \ll 1$) represents weak coupling strength.
	The Gaussian white noise terms in Eq. (\ref{eq:8}) are mutually independent and satisfy $\left<\nu_i(t)\right> = 0$ and $\left<\nu_i(t)\nu_j(\tau)\right> = \delta_{ij} \delta(t-\tau)$. The diffusive coupling $G_y(y_i,y_j)=\mu (y_j-y_i)$ is introduced only between the slow variables. Similarly to Eq. (\ref{eq:7}), the reduced phase equation for each oscillator is given by $\dot \theta_i=\omega_e+\sqrt{D_e}\xi_i(t)+\mu Z_y(\theta_i)G_y(\theta_i,\theta_j)$. By considering the phase difference $\phi = \theta_1 - \theta_2$, which is a slow variable, and applying the averaging procedure, we can derive the equation for $\phi$ as~\cite{Kuramoto1984}
	\begin{equation}
		\dot \phi=\sqrt{2D_e}\xi(t)+\mu \Gamma_d(\phi),
		\label{eq:9}
	\end{equation}
	where $\Gamma_d(\phi)=\Gamma(\phi)-\Gamma(-\phi)$ is the antisymmetric part of the phase coupling function $\Gamma(\phi)=\frac{1}{2\pi}\int_{0}^{2\pi} Z_y(\phi+\psi)G_y(\phi+\psi,\psi) \mathrm{d}\psi$. The solution of $\Gamma_d(\phi)=0$ represents a synchronized state of the two SISR oscillators, which is stable (unstable) when $\Gamma^\prime_d(\phi)<0$ [$\Gamma^\prime_d(\phi)>0$]. As the two oscillators are identical and the coupling is symmetric, the in-phase ($\phi=0$) and antiphase ($\phi=\pm \pi$) synchronized states are always the solutions as shown in Fig.~\ref{fig:4}(a). It is notable that both synchronized states are stable in the parameter regime considered here (although the stability of the antiphase synchronization is weaker).
	
	Since noise is present in our coupled system (\ref{eq:8}), the phase difference $\phi$ does not converge to a fixed value but forms a stationary distribution with peaks corresponding to the stable synchronized states \cite{Zhu2020ND}. This distribution depends on both the noise intensity and the coupling strength. We performed MC simulations of the coupled system (\ref{eq:8}) with initial phase differences uniformly distributed in $[-\pi,\pi]$. As shown in Fig.~\ref{fig:4}(b)-(c), the phase difference tends to localize around the in-phase and antiphase synchronized states. Increasing the coupling strength can enhance the localization and more clearly separate the two states, which shows the competing relationship between the coupling-induced synchronization and noise-induced desynchronization. The distributions of the phase difference obtained by MC simulations of the original system can be well reproduced by the reduced phase equation~(\ref{eq:9}) with the coupling function $\Gamma_d(\phi)$ obtained theoretically, wherein a higher peak is observed at the more stable in-phase state than at the less stable antiphase state. As expected, for smaller coupling strength, the phase-difference distribution is more accurately predicted.
	
	{\it Conclusion.-}
	We have investigated the phase dynamics of the SISR oscillator exhibiting noise-induced coherent oscillations. The transition positions on each branch were accurately obtained via DMC or FPTD, and an approximate hybrid system was established by connecting the dynamics on the two slow branches by discontinuous transitions. We performed phase reduction on the hybrid system and obtained the reduced phase equation by further incorporating the effective frequency and effective noise intensity. The reduced equations were applied to the analysis of a periodically forced SISR oscillator and a pair of mutually coupled identical SISR oscillators. The good agreement between the predicted dynamics and the results of the original model proved the accuracy and efficiency of our reduction method. The analysis in this Letter can be readily extended to more complex situations such as nonidentical coupled oscillations and networks. Also, more accurate results would be obtained by considering higher-order approximations \cite{Rosenblum2019Chaos,Leon2019PRE}. Moreover, despite that the considered SISR oscillator has only one-dimensional slow dynamics, the present approach can also be extended to systems with higher-dimensional slow dynamics as long as the oscillation is coherent. More details will be reported in our future works.
	
	{\it Acknowledgments.-}
	We thank the anonymous reviewers for valuable and insightful comments. J.Z. acknowledges support from JSPS KAKENHI JP20F40017, Natural Science Foundation of Jiangsu Province of China (Grant No. BK20190435), and the Fundamental Research Funds for the Central Universities (Grant No. 30920021112). Y.K. thanks JSPS KAKENHI JP20J13778 and JP22K14274 for financial support. H.N. thanks JSPS KAKENHI JP22K11919, JP22H00516, JPJSBP120202201, and JST CREST JP-MJCR1913 for financial support.
	
	

\newpage

\begin{center}
	{\large \bf Supplemental Material}\\
	{\bf Phase dynamics of noise-induced coherent oscillations in excitable systems}\\
	{\bf J. Zhu, Y. Kato and H. Nakao}
\end{center}

\section{Stochastic periodic orbits for the excitable FHN system}

For the timescale separation parameter considered in this paper, the stochastic orbits as shown in Fig.~1 in the main text are very coherent. We apply the distance matching condition (DMC), which we developed in \cite{Zhu2021PRRs}, to obtain the critical transition position on the left branch of the $x$ nullcline. The calculation of the transition position on the right branch is similar.

For each fixed value of the slow variable $y$ on the left branch, there is a corresponding mean first passage time (MFPT) $T_e(y)$ characterizing the difficulty of the transition from the left branch to the middle one. The MFPT $T_e(y)$ can be easily obtained from Eq.(1) in the main text describing the system as \cite{Gardiner1985s}
\begin{equation}
	T_{\rm e}(y)=\frac{2 \pi}{\sqrt{\lvert U''(x_m) \rvert U''(x_l)}}{\rm exp}\left(\frac{2\left(U(x_m)-U(x_l)\right)}{D_\nu}\right),
	\label{eq:A1}
\end{equation}
where $U$ represents the potential function of the deterministic fast subsystem of system (1) in the main text, $x_l$ and $x_m$ are values of $x$ on the left (stable) and middle (unstable) branches for the fixed slow variable $y$, respectively, and $D_\nu$ is the noise intensity. The mean first passage velocity (MFPV) can be accordingly defined as
\begin{equation}
	V_{\rm e}(y)=\frac{S(y)}{T_{\rm e}(y)},
	\label{eq:A2}
\end{equation}
where $S(y)$ is the distance between the left and middle branches, which is also a function of the slow variable $y$. By integration with respect to time using $\mathrm{d}y = \varepsilon( x + a ) \mathrm{d}t$ and $x = f_l^{-1}(y)$ ($f(x)=x-\frac{x^3}{3}$ and subscript $l$ denotes the solution on the left branch) and substituting Eq.(\ref{eq:A1}) into Eq.(\ref{eq:A2}), the DMC is expressed as
\begin{equation}
	\frac{1}{2 \pi \varepsilon }\int^{y_{l}}_{y_{0}} \frac{S(y)\sqrt{\lvert U''(x_m) \rvert U''(x_l)}}{\left(f_l^{-1}(y)+a\right) \, {\rm exp}\left(\frac{2\left(U(x_m)-U(x_l)\right)}{D_\nu}\right)}\,dy=S(y_{l}),
	\label{eq:A3}
\end{equation}
where $y_{0}$ is the starting position which can be chosen arbitrarily above but not close to the final transition position $y_{l}$ (the result depends on $y_0$ only slightly).

\begin{figure*}
	\centering
	\includegraphics[width=0.48\textwidth]{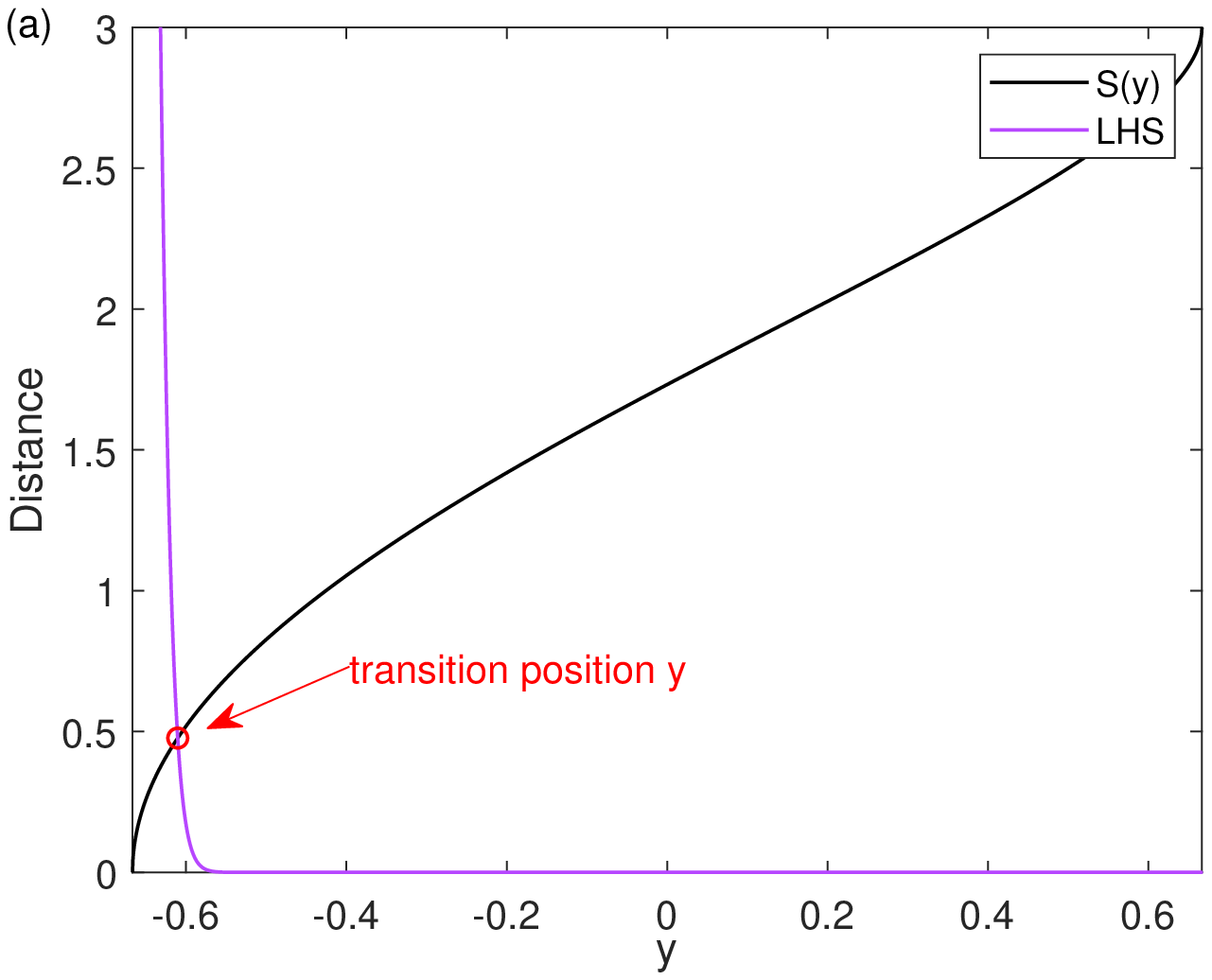}
	\includegraphics[width=0.48\textwidth]{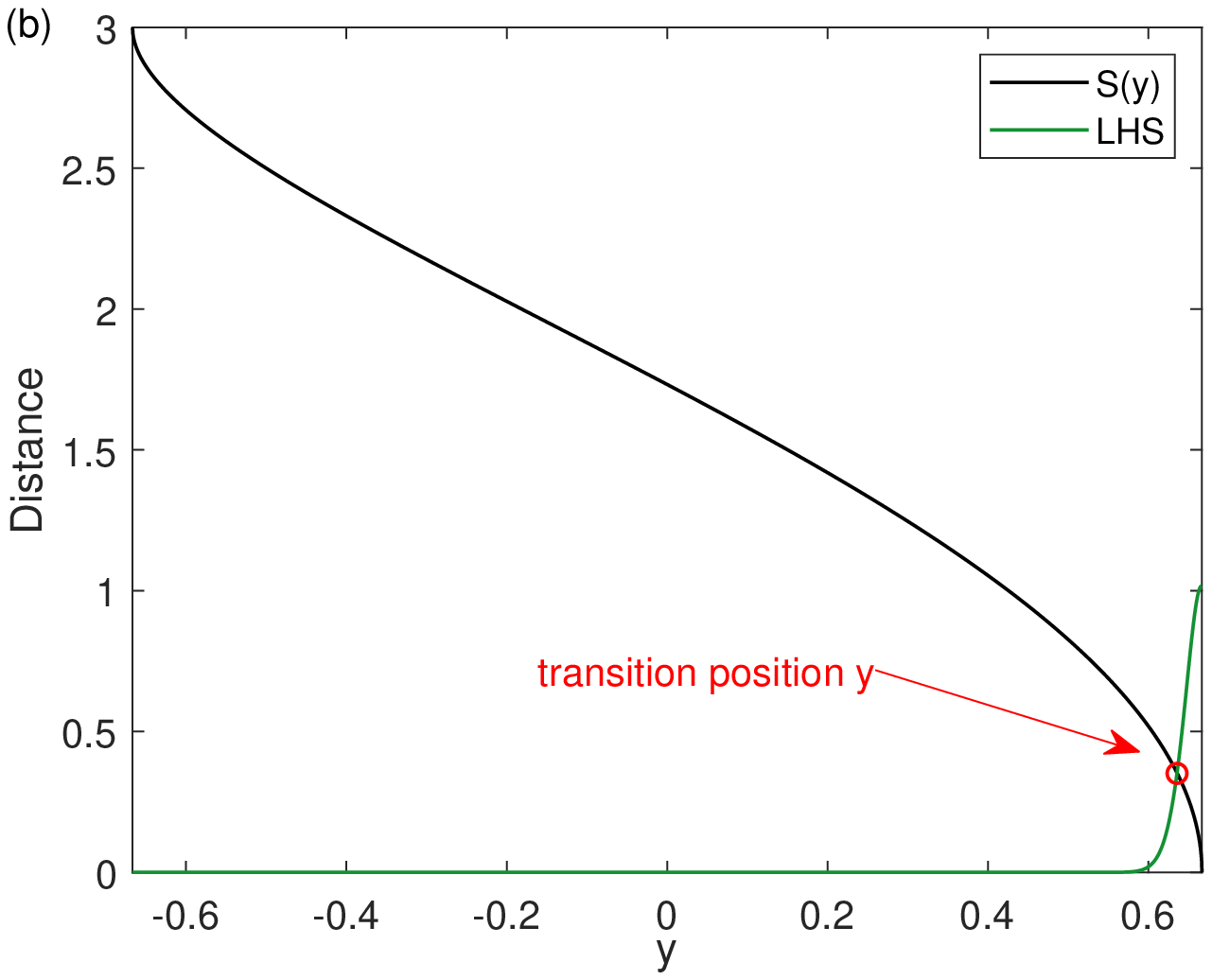}
	\caption{\label{fig:S0}Transition positions (red circles) on the left and right branches predicted by DMC (Eq.(\ref{eq:A3})). (a) Left branch. (b) Right branch.}
\end{figure*}

By numerically evaluating the left-hand side(LHS) and right-hand side(RHS) of Eq.(\ref{eq:A3}), the intersection point of these two curves gives the transition position. For the parameters considered in this paper, the results are shown in Fig.~\ref{fig:S0}. It is found that the transition happens in a narrow range of $y$ and the LHS remains nearly zero initially for large $y$, which implies that the choice of $y_0$ is not important as long as it is not too close to the transition position.

After determining the transition positions on the left ($y_l$) and right ($y_r$) branches, the stochastic periodic orbit can be obtained by connecting the slow dynamics along the $x$ nullcline with the jump processes at the transition positions (see Fig.~1 in the main text). The period can thus be approximated as
\begin{equation}
	T_h=\int_{y_r}^{y_l} \frac{dy}{\varepsilon \left(f_l^{-1}(y)+a\right)}+\int_{y_l}^{y_r} \frac{dy}{\varepsilon \left(f_r^{-1}(y)+a\right)}.
	\label{eq:A4}
\end{equation}
Therefore, the frequency of the hybrid system can be obtained as $\omega_h=2\pi T_h^{-1}$, which approximates the average frequency of the SISR oscillator.

\begin{figure}
	\centering
	\includegraphics[width=0.48\textwidth]{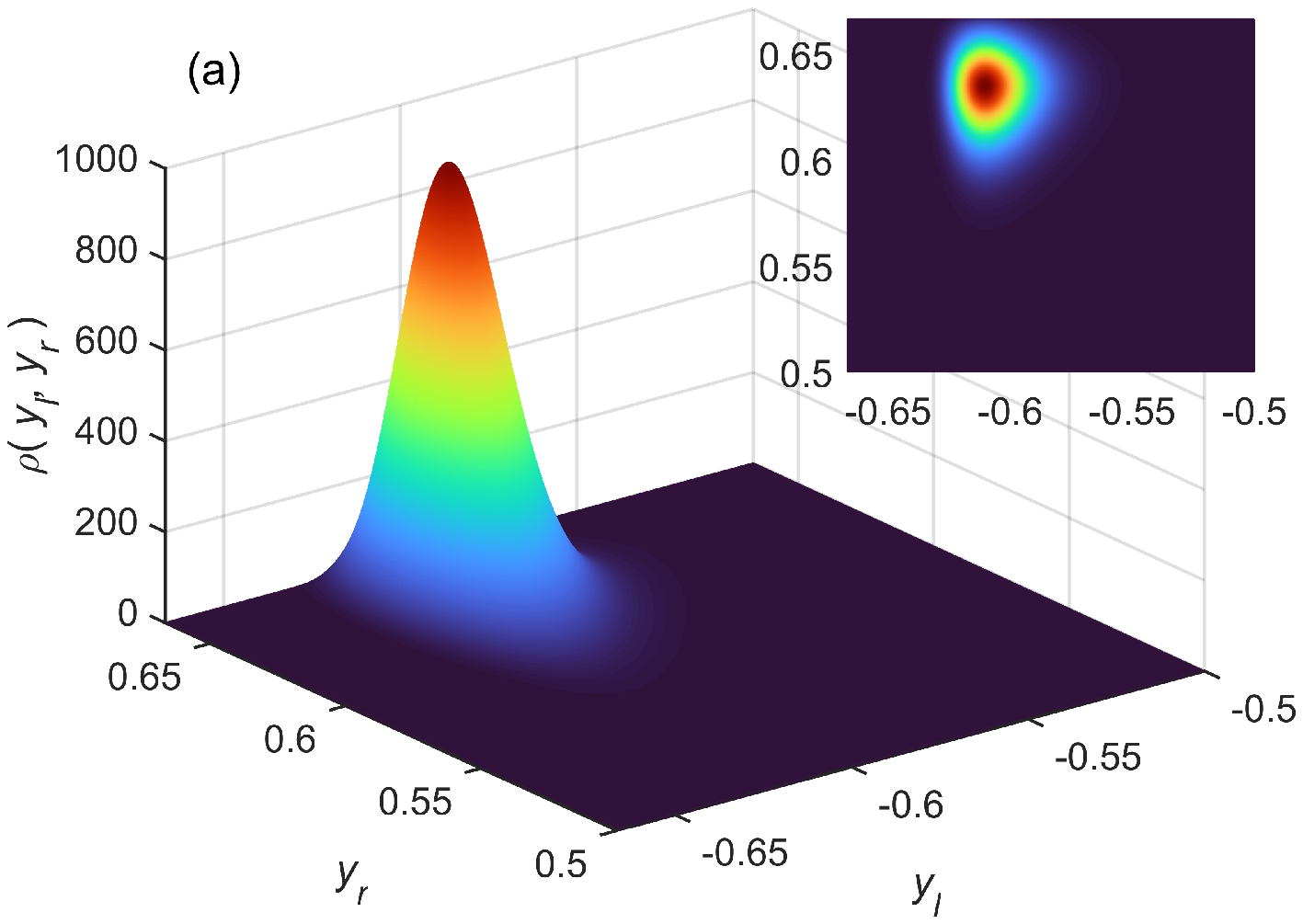}
	\includegraphics[width=0.48\textwidth]{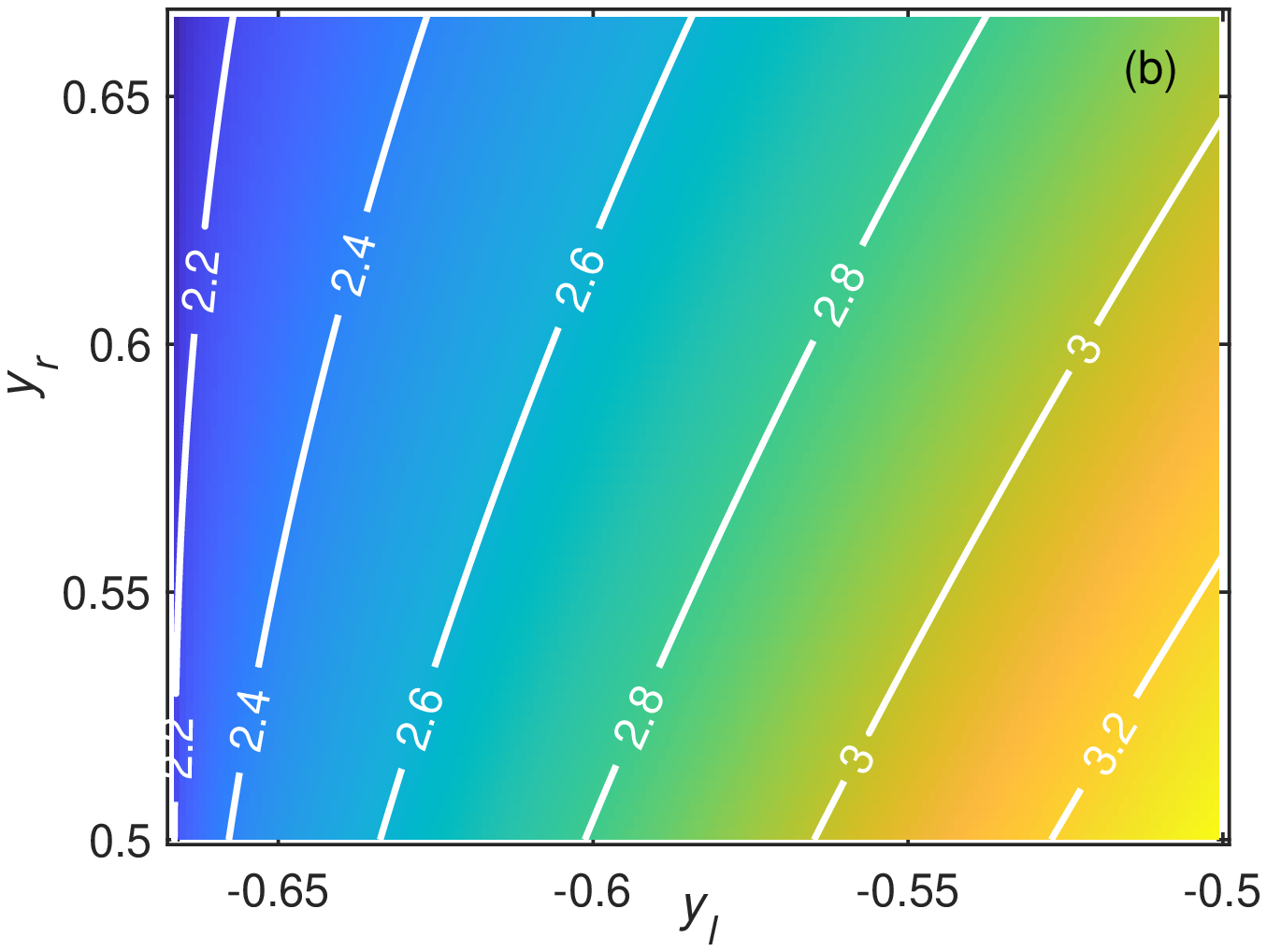}
	\caption{\label{fig:S1}(a) Joint probability density for the transition positions. The inset shows the mechanism of coherent behavior that the transition positions accumulate in a small range. (b) Oscillation frequency $\omega_h$ versus different transition positions on the left branch ($y_l$) and right branch ($y_r$).}
\end{figure}

To estimate the dispersion of the stochastic periodic orbit, we can consider the first passage time distribution (FPTD) \cite{Gardiner1985s,Lim2010JCNs,Li2019PREs}. For fixed $y$, the survival probability that the state remains on the LHS of the middle branch can be defined as
\begin{equation}
	G(x,t)=\text{Prob}(T\geq t)=\int_{x}^{x_m} p(x^\prime,t|x,0) \mathrm{d}x^\prime=\int_{-\infty}^{x_m} p(x^\prime,t) \mathrm{d}x^\prime=G(t),
	\label{eq:A5}
\end{equation}
where $T$ is the first passage time for the state $x$ escaping from the middle branch $x_m$ for fixed $y$.	The derivative of the escape probability $1-G(t)$ with respect to $t$ gives the FPTD $\rho_l(t)=-\dot G(t)$. Thus, the escape rate $\lambda(t)$ satisfies:
\begin{equation}
	\lambda(t)=\frac{-\dot G(t)}{G(t)}.
	\label{eq:A6}
\end{equation}
Note that the escape rate is just the reciprocal of the MFPT, so that the survival probability can be easily solved as
\begin{equation}
	G(t)=\exp\left({-\int_{-\infty}^{t}\frac{1}{T_e(t^\prime)}\mathrm{d}t^\prime}\right).
	\label{eq:A7}
\end{equation}
Therefore, we can obtain the FPTD $\rho_l(t)$ as
\begin{equation}
	\rho_l(t)=\frac{1}{T_e(t)}\exp\left({-\int_{-\infty}^{t}\frac{1}{T_e(t^\prime)}\mathrm{d}t^\prime}\right).
	\label{eq:A8}
\end{equation}
Replacing $t$ with $y$ and by noting $\rho_l(t)=\rho_l(y)\left|\frac{\mathrm{d}y}{\mathrm{d}t}\right|$, we can finally achieve our derivation for the FPTD as a function of $y$:
\begin{equation}
	\rho_l(y)=-\frac{1}{\varepsilon (f_l^{-1}(y)+a) T_e(y)}\exp\left({-\int^{y}\frac{1}{\varepsilon (f_l^{-1}(y^\prime)+a) T_e(y^\prime)}\mathrm{d}y^\prime}\right).
	\label{eq:A9}
\end{equation}
The FPTD on the right branch is similar:
\begin{equation}
	\rho_r(y)=\frac{1}{\varepsilon (f_r^{-1}(y)+a) T_e(y)}\exp\left({-\int^{y}\frac{1}{\varepsilon (f_r^{-1}(y^\prime)+a) T_e(y^\prime)}\mathrm{d}y^\prime}\right).
	\label{eq:A10}
\end{equation}
The FPTDs on the left and right branches are shown in Fig.~1 in the main text, which are consistent with the results by Monte Carlo simulations. Assuming that the transition positions on the left and right branches are independent, the joint probability density of the transition positions is given by
\begin{equation}
	\rho(y_l,y_r)=\rho_l(y_l)\rho_r(y_r).
	\label{eq:A11}
\end{equation}
The joint probability density $\rho(y_l,y_r)$ shown in Fig.~\ref{fig:S1}(a) exhibits a clear peak at the transition positions, which is consistent with the coherent oscillations. Using Eq.(\ref{eq:A4}), the oscillation frequency $\omega_h(y_l,y_r)$ with different transition positions can be calculated as in Fig.~\ref{fig:S1}(b). Finally, we can approximate the mean and variance of $\omega_h$ as
\begin{equation}
	\begin{split}
		\left< \omega_h \right> &= \int\int \omega_h(y_l,y_r) \rho(y_l,y_r) \mathrm{d}y_l \mathrm{d}y_r,\\
		\sigma_{\omega_h} &= \int\int \omega_h(y_l,y_r)^2 \rho(y_l,y_r) \mathrm{d}y_l \mathrm{d}y_r-\left< \omega_h \right>^2.
	\end{split}
	\label{eq:A12}
\end{equation}
The effective noise intensity $D_e$ can thus be approximated as
\begin{equation}
	D_e=\frac{2\pi \sigma_{\omega_h}}{\left< \omega_h \right>}.
	\label{eq:A13}
\end{equation}
Therefore, via numerical calculation of the above equations for the parameter values used in the main text, the effective frequency and effective noise intensity can be estimated as $\omega_e\approx 2.5530$ and $D_e\approx 0.0095$ ($=\tilde{D_e}$ in the main text), which are close to the evaluated values via Monte Carlo simulations in the main text. 

The small errors in the estimation of $\omega_e$ and $D_e$ can be explained as follows. It can be observed that nearly all stochastic trajectories on the left branch undergo transitions before the tip of the $x$ nullcline, while the transitions on the right branch can happen after the tip with a finite probability (see the inset of Fig.~\ref{fig:S1}(a) for a blowup where the upper part is slightly truncated). This truncated probability can be quantitatively measured by integrating the joint probability density $\rho(y_l,y_r)$ within the considered area, which gives $0.9755 (\neq1)$. Although the distribution above the tip of the $x$ nullcline cannot be calculated within the framework of FPTD, it can be inferred that the lack of those trajectories will make the estimated effective frequency larger and effective noise intensity smaller, which contributes to the errors in our estimation.

\section{Phase reduction of the hybrid system}
We consider a hybrid system defined as follows (Eq.(4) in the main text):
\begin{equation}
	\begin{split}
		&\dot {\bm X} = {\bm F}({\bm X}), \text{if} ~{\bm X}\notin{\bm \Pi_i},\\
		&{\bm X}(t+0) = {\bm \Phi_i}({\bm X}(t)), \text{if} ~{\bm X}\in{\bm \Pi_i}, i=l,r,
	\end{split}
	\label{eq:B1}
\end{equation}
where ${\bm \Phi_i}({\bm X}(t))$ represents the transition function and ${\bm \Pi_i}=\left\{{\bm X}|L({\bm X})=y_{i}\right\}$ is the switching surface (for the left $(i=l)$ and right $(i=r)$ branches). Considering the large timescale separation, we assume that, when the state crosses the critical transition position, it will be instantly attracted to the other stable branch of the $x$ nullcline, so that $L({\bm X})=y$ and ${\bm \Phi_l}({\bm X}) = \left[2 \cos(\varphi), y\right]^\top, {\bm \Phi_r}({\bm X}) = \left[2 \cos\left(\varphi+\frac{2\pi}{3}\right), y\right]^\top$, where $\varphi=\frac{1}{3} \arccos\left(-\frac{3}{2}y\right)$ (solving the cubic equation $x-\frac{x^3}{3}-y=0$ by using the trigonometric functions). We assume that Eq.(\ref{eq:B1}) has a stable piecewise-continuous limit-cycle solution $\bm X_0(t)$, and introduce an asymptotic phase function $\Theta(\bm X)$ in its basin of attraction. We denote this limit cycle as $\gamma_s$ and introduce the phase variable of the system as $\theta(t) = \Theta(\bm X(t))$, which increases with a constant frequency $\omega$ in the absence of perturbations. The state of $\gamma_s$ is expressed as $\bm X_0(\theta)$ as a function of the phase. Applying the phase reduction method for hybrid systems on system (\ref{eq:B1}) under a weak perturbation $\bm P(\bm X,t)$, the reduced equation can be given as
\begin{equation}
	{\dot \theta(t)}=\omega+\frac{\partial \Theta({\bm X})}{\partial {\bm X}} \cdot \bm P(\bm X,t)
	\approx \omega+ \bm Z(\theta) \cdot \bm P\left(\theta,t\right),
	\label{eq:B2}
\end{equation}
where we have approximated $\bm X(t)$ by the state $\bm X_0(\theta(t))$ on $\gamma_s$ having the same phase value $\theta(t) = \Theta(\bm X(t))$ as $\bm X(t) $ and $\bm Z(\theta)=\frac{\partial \Theta({\bm X})}{\partial {\bm X}}\big|_{\bm X=\bm X_0(\theta)}$ is the phase sensitivity function of $\gamma_s$.

The phase sensitivity function can be obtained by solving the following adjoint system \cite{Shirasaka2017PREs} for hybrid limit-cycle oscillators:
\begin{equation}
	\begin{split}
		&\omega \frac{\mathrm{d}}{\mathrm{d}\theta}\bm Z(\theta)=-{\bf J}(\theta)^\top\bm Z(\theta), \text{if} ~\bm X(\theta) \notin \bm \Pi_i,\\
		&\bm Z\left(\theta(t)\right)={\bf C}_i^\top \bm Z\left(\theta(t+0)\right), \text{if} ~\bm X(\theta) \in \bm \Pi_i,
	\end{split}
	\label{eq:B3}
\end{equation}
with the normalization condition:
\begin{equation}
	\bm Z(\theta) \cdot \bm F(\bm X(\theta))=\omega.
	\label{eq:B4}
\end{equation}
Here, ${\bf J}(\theta)$ is the Jacobi matrix of $\bm F(\bm X)$ at $\bm X=\bm X_0(\theta)$ and the superscript $\top$ denotes transpose. The saltation matrix ${\bf C}_i$ describing the change in $\bm Z$ at the switching surface is given by \cite{Shirasaka2017PREs}:
\begin{equation}
	{\bf C}_i = D {\bm \Phi}_i ( \bm X_0(t_i) )- [ D {\bm \Phi}_i ( \bm X_0(t_i) ) \dot{\bm X}_0(t_i) - \dot{\bm X}_0(t_i + 0)  ] \otimes \left( \frac{\nabla L( {\bm X}_0(t_i) )}{\nabla L( {\bm X}_0(t_i) ) \cdot \dot{\bm X}_0(t_i)} \right),
	\label{eq:B6}
\end{equation}
where $D {\bm \Phi}_i$ is the Jacobi matrix of ${\bm \Phi}_i$ and $t_i$ denotes the transition time. The symbol $\otimes$ represents the Kronecker product. Through backward integration, the adjoint equation (\ref{eq:B3}) can be numerically solved \cite{Ermentrout2010Books,Ermentrout1996NCs}. For details of the phase reduction approach on the hybrid system, the readers are referred to Ref.~\cite{Shirasaka2017PREs}.

\section{Direct method for phase sensitivity function}

For deterministic systems with a stable limit cycle, a fixed perturbation at a fixed timing can produce a fixed change of the phase value. The phase response function characterizing the change of the phase value caused by a perturbation $\bm \delta$ given at the phase $\theta$ can be represented as
\begin{equation}
	g(\theta;\bm \delta)=\Theta(\bm X_0(\theta)+\bm \delta)-\Theta(\bm X_0(\theta))=\Theta(\bm X_0(\theta)+\bm \delta)-\theta.
	\label{eq:C1}
\end{equation}
By assuming the perturbation $\bm \delta$ to be small, the following Taylor expansion can be obtained:
\begin{equation}
	\Theta(\bm X_0(\theta)+\bm \delta)=	\Theta(\bm X_0(\theta))+\bm Z(\theta)\cdot \bm \delta+o(|\bm \delta|),
	\label{eq:C2}
\end{equation}
where $\bm Z(\theta)$ is the phase sensitivity function. Therefore, by applying a sufficiently small perturbation $\bm \delta=\delta \bm e_i$ ($\bm e_i$ is a unit vector with only a single nonzero value at its $i$-th component), the $i$-th component of the phase sensitivity function $Z_i(\theta)$ can be numerically measured as
\begin{equation}
	Z_i(\theta)=\lim\limits_{\delta \to 0}\frac{g(\theta;\delta \bm e_i)}{\delta}.
	\label{eq:C3}
\end{equation}
The phase value of $\Theta(\bm X_0(\theta)+\bm \delta)$ can be calculated by evolving the state $\bm X_0(\theta)+\bm \delta$ for several periods of the oscillator until it converges to the stable limit cycle. However, for the SISR oscillator, the phase value differs between realizations due to noise. We denote by $\bm Y(t)$ and $\bm X(t)$ the oscillator states at time $t$ with initial conditions $\bm Y(0)=\bm X_0(\theta)+\bm \delta$ and $\bm X(0)=\bm X_0(\theta)$. Then, according to Eq.(5) in the main text, the corresponding phase values evolve as
\begin{equation}
	\begin{split}
		\Theta(\bm Y(t))&=\Theta(\bm Y(0))+\omega_e t+\sqrt{D_e}W_1(t),\\
		\Theta(\bm X(t))&=\Theta(\bm X(0))+\omega_e t+\sqrt{D_e}W_2(t),
	\end{split}
	\label{eq:C4}
\end{equation}
where $W_1(t)$ and $W_2(t)$ are independent Wiener processes. From Eq.(\ref{eq:C2}), the following time-dependent stochastic phase response is obtained:
\begin{equation}
	g(\theta;\bm \delta,t)=\Theta(\bm Y(t))-\Theta(\bm X(t))=\bm Z(\theta)\cdot\bm \delta+\sqrt{2D_e}W(t)+o(|\bm \delta|),
	\label{eq:C5}
\end{equation}
where $W(t)$ is another Wiener process. By introducing a phase response function rescaled by the perturbation strength as $g_\delta(\theta;\delta \bm e_i,t)=\delta^{-1}g(\theta;\delta \bm e_i,t)$, the $i$-th component of $\bm Z(\theta)$ can be expressed as
\begin{equation}
	Z_i(\theta)= \lim\limits_{\delta \to 0}\left<g_\delta(\theta;\delta \bm e_i,t) \right>,
	\label{eq:C6}
\end{equation}
where we used $\left<W(t)\right>=0$. The standard deviation of $g_\delta(\theta;\delta \bm e_i,t)$ is $\sigma_g=\delta^{-1}\sqrt{2D_e t}$. Therefore, we can also measure the effective noise intensity from $\sigma_g$. It is interesting to see that there is a dilemma here: decreasing the perturbation strength $\delta$ can increase the linearity and thus the accuracy of the phase sensitivity function computed by the above direct method, making it closer to the theoretical result obtained via the adjoint method for the hybrid system; while it may also increase the standard deviation of the result, which may make the simulation results more stochastic. This dilemma can be observed in Fig.~2 in the main text.

\end{document}